\documentclass[12pt]{article}

\usepackage{amsthm,amsmath,amsfonts,amssymb}
\usepackage[authoryear]{natbib}
\usepackage[colorlinks,citecolor=blue,urlcolor=blue,backref=page,backref=page]{hyperref}
\usepackage{graphicx}

\usepackage[justification=raggedright]{subcaption}
\usepackage{xcolor}
\usepackage{afterpage}
\usepackage{placeins}
\usepackage{cancel}

\usepackage{ulem}    % For strike-through
\usepackage{comment} % For block commenting

\usepackage{tikz}
\usetikzlibrary{bayesnet}
\usepackage{booktabs}
\usepackage{fancyhdr} 
\usepackage[margin=1.2in]{geometry}

\newcommand{\keywords}[1]{\textbf{Keywords:} #1}

\newcommand{\mymodel}{FSLDS}

\newcommand{\disc}{s}

\theoremstyle{plain}

\theoremstyle{definition}

\theoremstyle{remark}

\title{Detecting State Changes in Functional Neuronal Connectivity using Factorial Switching Linear Dynamical Systems}

\author{
  Yiwei Gong\thanks{Dept. of Statistics and Data Sciences, UT Austin. \texttt{yiwei.gong@utexas.edu}} \and
  Susanna B. Mierau\thanks{Dept. of Neurology, BWH and Harvard Medical School. \texttt{smierau@bwh.harvard.edu}} \and
  Sinead A. Williamson\thanks{Apple MLR. \texttt{sa\_williamson@apple.com}}
}

\date{Last updated: \today}

\begin{document}
\maketitle

\begin{abstract}
A key question in brain sciences is how to identify time-evolving functional connectivity, such as that obtained from recordings of neuronal activity over time. We wish to explain the observed phenomena in terms of latent states which, in the case of neuronal activity, might correspond to subnetworks of neurons within a brain or organoid. Many existing approaches assume that only one latent state can be active at a time, in contrast to our domain knowledge. We propose a switching dynamical system based on the factorial hidden Markov model. Unlike existing approaches, our model acknowledges that neuronal activity can be caused by multiple subnetworks, which may be activated either jointly or independently. A change in one part of the network does not mean that the entire connectivity pattern will change. We pair our model with scalable variational inference algorithm, using a concrete relaxation of the underlying factorial hidden Markov model, to effectively infer the latent states and model parameters. We show that our algorithm can recover ground-truth structure and yield insights about the maturation of neuronal activity in microelectrode array recordings from in vitro neuronal cultures.
\end{abstract}


\section{Introduction}

The brain receives information from the surrounding world before reacting correspondingly and systematically. As a complex system, it acts differently across time and locations. Thus, analytical tools that can detect correlations in spatial and temporal activity are required to study the brain's activity. The dynamic networks that underlie these brain processes at the microscale can be studied using  microelectrode array (MEA) recordings of neural signals across a whole population of neural cells \citep{schroter2017micro}. % \cite{mea1}. 
While MEA recordings do not allow us to target specific neurons, neural signals, highly correlated in time, from neurons near multiple electrodes can reveal the spatial-temporal features of brain network function. Understanding this functional connectivity \textit{in vitro} is important for understanding microscale brain function and development \textit{in vivo}~\citep{humphries2017}.  

Previous studies of microscale functional neuronal networks have focused on the network topology \citep{downes2012emergence,schroeter2015emergence,sit2024mea}. These studies estimate a functional graph by considering pair-wise correlations between electrodes over a period of time. Graph theoretical metrics are then used to compare the topology of the functional networks including the spatial arrangement of nodes \citep{schroter2017micro}. These approaches have two important limitations. Firstly, they estimate a single pattern of functional connectivity based on minutes of MEA recording. This ignores the fact that brain networks are adaptable on much faster time-scales and precludes study of such short-term changes in connectivity. Secondly, they rely on pair-wise comparison of nodal activity to determine functional connectivity, which may not capture patterns of activity observed in subnetworks detected by more than two electrodes.

Other works have used discrete state-space models to capture evolving functional connectivity. These approaches capture the idea that a dynamical system---in our case, temporal MEA recordings---can be explained in terms of a small number of underlying behavioral patterns. In the simplest state-space model, hidden Markov models (HMMs) model neuronal activity as arising from one of a small number of underlying activity patterns \citep{paninski2010new}. Switching linear dynamical systems (SLDS) extend this by characterizing each activity pattern using a linear dynamical system \citep{petreska2011dynamical}. Extensions add additional structure between the latent states, allowing for more flexible dynamics \citep{pmlr-v54-linderman17a,nassar2018tree,geadahparsing, fox2008nonparametric}. These approaches, however, make the unrealistic assumption that neuronal activity can be partitioned into a small number of \textit{a priori} independent activity patterns. They lack the capacity to detect multiple network patterns simultaneously, as occurs in complex brain networks. More broadly, this single-state bottleneck is a fundamental limitation whenever the data-generating process involves multiple latent components whose activations overlap in time.

 New methods are needed to identify changes in functional connectivity across seconds to minutes during MEA recordings, without making unrealistic assumptions about the nature of connectivity changes. In this work,  we posit that neuronal activity can be decomposed into multiple subnetworks that can be activated jointly or separately. This allows us to capture overlapping patterns of behavior without the need to create a single state for each combination of active subnetworks. 
 
The starting point for our approach is the factorial hidden Markov model \citep[FHMM,][]{ghahramani1995factorial}. The FHMM is an extension of the hidden Markov model, where multiple latent Markov chains evolve independently. At any given point in time, the observation model is parametrized by the combination of latent states from all chains. In principle, using FHMMs to control the dynamics of multiple switching linear dynamical systems would allow us to capture the posited decomposition. However, the FHMM does not offer a \textit{practical} modeling solution, due to the challenge of performing inference in a discrete, combinatorically large latent space.

To solve this challenge, we begin by introducing the relaxed factorial switching model (RFSM), a continuous relaxation of the FHMM. Like the FHMM, the RFSM  allows multiple model components to be turned on and off independently. However, unlike the FHMM, by replacing the discrete latent chains with continuous variables via the concrete relaxation \citep{concrete}, the RFSM reduces the cost of inference from exponential to linear in the number of chains, and enables efficient gradient-based variational inference \citep{aevb}. The transition dynamics in the RFSM are parameterized using the output of a neural network. The input to this network can either be the previous state, yielding a Markovian model analogous to the FHMM, or a representation of all previous states, offering more flexible dynamics when the data exhibits steoreotypy or periodicity. This non-Markovian extension goes beyond the standard FHMM and is made straightforward by the continuous formulation.

We use the RFSM as the backbone of our proposed Factorial Switching Linear Dynamical System (\mymodel{}). The latent features in the RFSM correspond to subnetworks of neurons, which can be activated independently. When active, each subnetwork's activity is governed by a linear dynamical system. This yields an interpretable model that we use to explore latent structure in multivariate time-series data. We demonstrate this by exploring the latent structure of MEA recordings, showing that we can recover known state changes and revealing a developmental increase in the number of latent features (or subnetworks) in human induced pluripotent stem cell (iPSC)-derived neuronal networks in vitro.

To summarize our contributions:
\begin{enumerate}
    \item We propose \mymodel{}, a flexible, interpretable model for multivariate time-series data. The transition dynamics of \mymodel{} are governed by a neural network, allowing us to model either Markovian or non-Markovian state changes (Sections~\ref{sec:concreteFHMM} and \ref{sec:our_model}). 
    \item The continuous latent structure of \mymodel{} facilitates gradient-based inference using autoencoded variational Bayes (Section~\ref{sec:inference}). This makes \mymodel{} a more tractable alternative to the FHMM.
    \item We show that \mymodel{} can identify interpretable subnetwork activity both murine and human based in-vitro neuronal cultures (Section~\ref{sec:results}).
\end{enumerate}

\section{Background}
Our model is a factorial state-space model that extends existing time-series models by allowing multiple latent factors to be active at once, using a continuous relaxation of a factorial hidden Markov model \citep[FHMM,][]{ghahramani1995factorial} to control which combination of latent factors are active. This leads to a more interpretable model for temporally evolving neuronal functional connectivity, where we assume that observations are generated from an additive combination of multiple subnetworks, that may not all activate synchronously. Optionally, we relax the Markovian assumption to allow more flexible dynamics such as periodicity.

In Section~\ref{sec:bg_ssms}, we review existing work in state space models for time series data, with a particular focus on models appropriate for modeling MEA recordings. In Section~\ref{sec:fhmm}, we focus specifically on factorial state space models, particularly the FHMM which forms the backbone of our approach. We then discuss Bayesian inference in complex discrete state-space models in Section~\ref{sec:bg_variational}.

\subsection{Modeling functional connectivity using discrete state space models}\label{sec:bg_ssms}
Discrete state-space models such as hidden Markov models ~\citep[HMMs,][]{baum1966statistical} provide an interpretable model for time-series data, where we assume that each time-point belongs to one of a finite number of states. Observations are assumed iid given a state-specific emission-distribution. HMMs have been used to uncover latent structure in neuronal activity \citep{camproux1996hidden,xydas2011revealing,li2018spike}. Switching dynamical systems replace the emission model with a dynamical system. The simplest example of this is a switching linear dynamical system ~\citep[SLDS,][]{ackerson1970state}, where a Markovian discrete state process switches between autoregressive models with state-specific parameters, which can lead to a more flexible model for MEA recordings \citep{petreska2011dynamical}. Building on SLDS, more sophisticated models have been developed to handle more complex scenarios, such as recurrent switching linear dynamical system ~\citep[rSLDS,][]{pmlr-v54-linderman17a} and tree-structured rSLDS~\cite{nassar2018tree}. These models introduce recurrent dependency structures, allowing memory between consecutive time points.  Additionally, when linear assumptions prove insufficient, nonlinear dynamical system methods have been proposed~\cite{dong2020collapsed,smith2021reverse}.~\cite{ojeda2021switching} introduces a variational RNN model with attention to incorporate both nonlinear dynamics and non-Markovian switches. %Nonlinear RNN and SLDS based on the Jacobian of the RNN could be co-trained to eliminate extra postprocessing steps and to increase more flexibility when approximating dynamics~\cite{smith2021reverse}. %However, these models often focus more on the amplitudes of node activities, without explaining how the network's connectivity evolves over time. In addition, single-HMM based methods assume all nodes change to the same state simultaneuously, making it difficult to capture subgraph information of nodes or to learn the co-activation of nodes.

%To address these limitations, we propose a recurrent switching nonlinear dynamical system with a factorial latent discrete structure. We also design a scalable variational framework to learn the model, aiming to describe temporally modulated interactions between nodes while simultaneously learning changes in amplitudes.
\subsubsection{Factorially structured discrete state space models}\label{sec:fhmm}

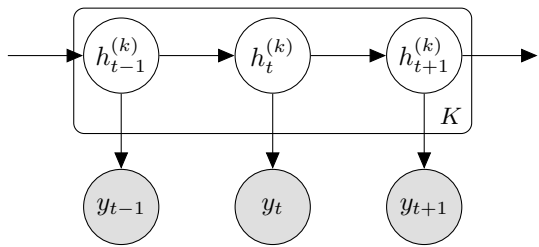
\begin{figure}[ht]
    \centering
    \begin{tikzpicture}

  % Define nodes
  \node[const] (h0){};%
  \node[latent, right=of h0,minimum size=1cm] (h1) {$\disc^{(k)}_{t-1}$};%
  \node[latent, right=of h1,minimum size=1cm] (h2) {$\disc^{(k)}_{t}$};%
  \node[latent, right=of h2,minimum size=1cm] (h3) {$\disc^{(k)}_{t+1}$};%
  \node[const, right=of h3,minimum size=1cm] (h4){}; %
  \node[obs, below=of h1,minimum size=1cm]                               (y1) {$y_{t-1}$};%
  \node[obs, below=of h2,minimum size=1cm]                               (y2) {$y_{t}$};%
  \node[obs, below=of h3,minimum size=1cm]                               (y3) {$y_{t+1}$};
  % Connect the nodes
  \edge {h1} {y1} ; %
  \edge {h2} {y2} ; %
  \edge {h3} {y3} ; %
  \edge {h1} {h2} ; %
  \edge {h2} {h3} ; %
  \edge {h0} {h1} ;
  \edge {h3} {h4};

  % Plates
  \plate {p1} {(h1)(h2)(h3)} {$K$} ;

\end{tikzpicture}
    \caption{Graphical model for a factorial hidden Markov models. $K$ discrete Markovian processes jointly influence the observation $y_t$. By contrast, a standard hidden Markov model is controlled by a single discrete process.}
    \label{fig:fhmm}
\end{figure}

The above models assume the underlying behavior switches between a small set of patterns. This may be unrealistic in complex time series such as MEA recordings. 

In practice, two timepoints may share some characteristics, but not others. This observation motivates the factorial hidden Markov model \citep[FHMM,][]{ghahramani1995factorial}, where multiple hidden factors $\{\disc_t^{(k)}\}_{k=1}^K$ concurrently influence the observed dynamics at each time point $y_t$.
\begin{equation}
\begin{aligned}
    \disc_t^{(k)} | \disc_{t-1}^{(k)} \sim& \text{Categorical}\left(\pi_{\disc_{t-1}^{(k)}}^{(k)}\right)\\
    y_{t}|\disc_t^{(1)},\dots, \disc_t^{(K)} \sim& f_y\left(\theta_{\disc_t^{(1)}}, \dots, \theta_{\disc_t^{(K)}}\right).
    \end{aligned}\label{eqn:fhmm}
\end{equation}
Each latent factor $\disc_t^{(k)}$ evolves according to an independent Markovian process with transition probabilities $\pi^{(k)}$. The emission process at time $t$ depends on the states of all $K$ latent factors, as depicted in Figure~\ref{fig:fhmm}. Typically, $f_{y}(\cdot)$ is an additive model that combines state-specific features $\theta_{\disc_t^{(k)}}$.

This factorial structure has previously been used in the context of dynamical systems. \citet{quinn2008factorial} assume a known switching pattern and use a FHMM to additively modify a single shared dynamical system. \citet{mudrik2024decomposed} use a similar factorial decomposition to combine dynamical systems; however, they use a sparse coding approach rather than explicitly incorporating temporal dynamics in the latent factors. These works focus on understanding and disentangling latent dynamics. Conversely, our approach disentangles the observation space, explaining observations in terms of multiple, additively combining subnetworks.

\subsection{Bayesian inference for discrete state space models}\label{sec:bg_variational}
The posterior distributions over the parameters of discrete state space models are often learned through Markov chain Monte Carlo (MCMC) using forward filtering backward sampling~\citep{west2013bayesian, godsill2004monte,olsson2011rao}. However, as model complexity increases, MCMC methods can suffer from slow mixing and high computational cost. Recent advances in variational inference provide a more computationally efficient alternative. 

Auto-Encoding Variational Bayes ~\citep[AEVB,][]{aevb} offers a flexible ``black box'' variational inference framework by using neural networks to specify a variational approximation and using gradient descent to optimize a sample-based estimate of the ELBO. However, gradient-based optimization methods cannot directly learn discrete variables, such as those present in switching dynamical systems. The REINFORCE algorithm \citep{weber2015reinforced} uses a Monte Carlo estimate of the gradient; however this estimate suffers from high variance~\citep{paisley2012variational}. Other approaches combine variational inference with sequential Monte Carlo \citep{maddison2017filtering,naesseth2018variational,le2017auto,lawson2022sixo}.

An alternative approach is to sidestep the use of discrete distributions altogether. The Concrete, or Gumbel-Softmax, distribution \citep{jang2016categorical,concrete},
\begin{equation}
    \text{Concrete}(x;\alpha, \phi)= (K-1)!\phi^{K-1}\prod_{k=1}^K\left(\frac{\alpha_k x_k^{-\phi-1}}{\sum_{j=1}^K \alpha_j x_j^{-\phi}}\right)\label{eqn:concrete}
\end{equation}
is a continuous distribution over the probability simplex that can be used as an approximation to a Categorical($\alpha_1,\dots, \alpha_K$) distribution. The location parameters $\alpha_1,\dots, \alpha_K$ in Equation~\ref{eqn:concrete} directly correspond to the probabilities in a Categorical distribution. The temperature $\phi\geq0$ controls the sparsity of samples. When $\phi=0$, we recover the Categorical($\alpha_1,\dots, \alpha_K$), with samples lying at the vertices of the simplex. As $\phi$ increases, the probability mass shifts towards the center of the simplex. Samples no longer put all their probability mass on a single category.  Instead, they assign probability mass to all categories,  with samples moving (in expectation) closer to the mean of the Categorical distribution as $\phi$ increases. 
Since the Concrete($\alpha, \phi$) distribution is continuous, we can use it directly in variational inference as an alternative to the Categorical distribution.

\section{A factorial switching linear dynamical system for 
modeling functional connectivity from MEA recordings}
We propose factorial switching linear dynamical systems (\mymodel{}), a factorially switching linear dynamical model appropriate for capturing functional connectivity patterns. We begin, in Section~\ref{sec:concreteFHMM}, by introducing a continuous relaxation of the FHMM, which serves as the backbone of our approach, which we present in Section~\ref{sec:our_model}. Replacing the discrete FHMM with a continuous relaxation allows us to perform variational inference without the need to explicitly enumerate over combinations of latent factors, as we describe in Section~\ref{sec:inference}.

\subsection{Relaxed factorial switching model}\label{sec:concreteFHMM}
The factorial nature of the FHMM leads to an exponentially growing parameter space. Exploration of this discrete state space is challenging, and traditional MCMC methods are prone to difficult and slow mixing~\citep{ghahramani1995factorial}. As we discussed in Section~\ref{sec:bg_variational}, variational inference provides an appealing alternative to MCMC, allowing us to exploit optimization techniques to learn an approximate posterior. However, as we saw, a naive implementation of variational inference methods does not allow for discrete latent states, since gradient descent is inherently designed for continuous variables.

To overcome this limitation, we introduce a continuous variant of the FHMM, that we refer to as a relaxed factorial switching model (RFSM). We relax the discrete latent state indicators $\disc_t$ in Equation~\ref{eqn:fhmm} into continuous variables, whose distribution is described using a Concrete distribution (Equation~\ref{eqn:concrete}). These continuous variables can be seen as indicators of the degree of contribution of the $k$th latent factor $\theta_k$. This provides significant computational benefits, as it enables the use of gradient-based optimization techniques that are more scalable and efficient than sampling-based methods.

In a FHMM, each latent factor is associated with a transition matrix that describes the probability of transitioning between states. In the RFSM, we no longer have discrete state indicators (as $\disc_t^{(k)}$ is continuous), so it no longer makes sense to associate a transition distribution with each latent state. Instead, following~\citet{becker2019switching}, we parametrize the Concrete distribution using the output of a neural network. 

We have significant flexibility in choosing this neural network. A simple choice here is to use a multi-layer perceptron (MLP) that takes the previous state $\disc_{t-1}$ as input, i.e., 

\begin{equation}
\label{eqn:mlp}
\begin{aligned}
    \alpha_{p, t}^{(k)}:=& \, \text{MLP}_p^{(k)}\left(\disc_{t-1}^{(k)}\right) \\
    \disc_{t}^{(k)}\sim & \, \text{Concrete}\left(\alpha_{p, t}^{(k)}, \phi\right)\\
    y_{t}|\disc_t^{(1)},\dots, \disc_t^{(K)} \sim& \, f_y\left(\sum_{k=1}^K \theta_k \disc_t^{(k)}\right),
    \end{aligned}
\end{equation}

where $f_y$  is an arbitrary emission model, as in \eqref{eqn:fhmm}. This parametrization, which we refer to as MLP-RFSM, retains the Markovian nature of the FHMM while relaxing it's discrete structure.

We parametrize the Concrete distribution with the output of a neural network. A simple choice here is to use a multi-layer perceptron (MLP) that takes the previous state $\disc_{t-1}$ as input, i.e.,

Alternative choices allow us to relax the Markovian assumption in the original FHMM, allowing the parameter of the concrete distribution to depend on the full history of the process. A lightweight way to do this is to use a recurrent neural network (RNN), i.e., 

\begin{equation}
\label{eqn:rnn}
\begin{aligned}
    \alpha_{p, t}^{(k)}, h_{p, t}^{(k)}:=& \, \text{RNN}_p^{(k)}\left(\disc_{t-1}^{(k)}, h_{p, t-1}^{(k)}\right) \\
\disc_{t}^{(k)}\sim & \, \text{Concrete}\left(\alpha_{p, t}^{(k)}, \phi\right)\\
    y_{t}|\disc_t^{(1)},\dots, \disc_t^{(K)} \sim& \, f_y\left(\sum_{k=1}^K \theta_k \disc_t^{(k)}\right),
    \end{aligned}
\end{equation}

where $h_{p, t-1}^{(k)}$ is the hidden state of the RNN. This parametrization, which we refer to as RNN-RFSM, is appropriate where we expect longer-term dependencies, such as periodicity or sterotypy. In our experiments, we explore both formulations of RFSM. In datasets where we observe stereotypy (based on Fourier analysis\footnote{We estimate the power spectral density (PSD) for each electrode using the periodogram method and identified the dominant oscillatory frequency as the PSD-maximizing frequency. The corresponding period, obtained by reciprocal transformation and averaged across electrodes, provides a frequency-domain estimate of periodicity.}), we use the RNN-RFSM formulation; in datasets with no obvious stereotypy, we use the simpler MLP-RFSM.

The computational advantage of using a continuous relaxation comes with the trade-off of losing directly interpretable transition probabilities that are inherent in discrete Markov processes. However, setting an appropriately low temperature $\phi$ means that in practice, we tend to have values that are very close to zero or one, as we see in Section~\ref{sec:results}. As we will see, this allows us to retain some level of interpretability, bridging the gap between computational efficiency and the practical understanding of the model's behaviour.

\subsection{Factorial switching linear dynamical systems}\label{sec:our_model} %AutoRegressive-Factorial Hidden Markov Model
As we have seen, HMMs, SLDSs, and their variants assume each time point corresponds to a single latent state. In practice, this might be unrealistic. For example, in developing brains, we expect new subnetworks to appear in addition to previous structure, rather than fully replacing the previous connectivity pattern. Further, we do not expect a subnetwork to have constant activity---instead, we want to disentangle changes in activity amplitude from changes in neuron connectivity. To achieve this, we first specify a set of $K$ features $\theta_k\in\mathbb{R}_+^M$, where $\theta_{k,m}$ represents the relative contribution\footnote{Since we scale $\theta_k$ by a continuous rate, $\theta_{k,m}$ can only be interpreted as relative contribution of the $m$th node relative to other nodes.} of the $m$th node to the $k$th subnetwork. We then use an RFSM to determine whether that feature is active at a given time, and model the value of the state when it is active using a linear dynamical system. The generative model is

\begin{equation}\begin{aligned}
    y_{t, m} &\sim\text{Poisson}\left(\sum_{k=0}^K \theta_{k,m} \disc_t^{(k)} \exp\left\{z_t^{(k)}\right\}\right),\; &m=1,\dots, M\\
    \disc_t^{(0)} &= 1,\\
    \disc_{t}^{(k)} &\sim \text{Concrete}\left(\alpha_{p,t}^{(k)},\phi\right),
    &k=1,\dots, K\\
    z_{t}^{(k)}\mid z_{t-1}^{(k)} &\sim \text{Normal}\left(A^{(k)} z_{t-1}^{(k)}, \,\sigma_p^{(k)2}\right), &k=0,\dots, K.\\
    \label{eqn:our_model}
\end{aligned}\end{equation}
where $\alpha_{p, t}^{(k)}$ is the output of either a MLP (as in Equation~\ref{eqn:mlp}) for the MLP-RFSM variant, or an RNN (as in Equation~\ref{eqn:rnn}) for the RNN-RFSM variant.

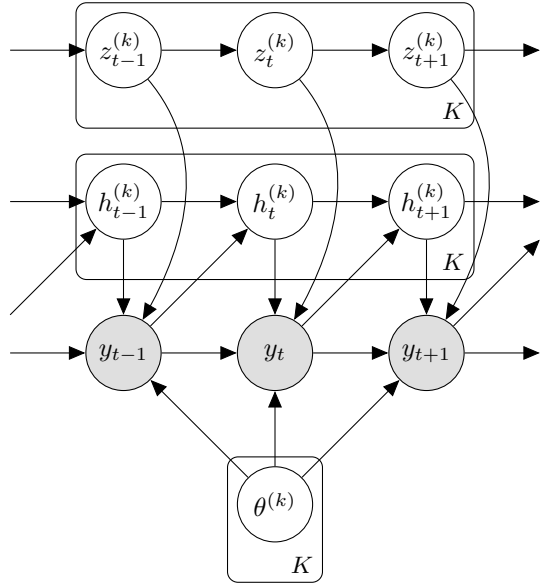
\begin{figure}[ht]
    \centering
\begin{tikzpicture}

  % Define nodes
  \node[const] (h0){};%
  \node[latent, right=of h0,minimum size=1cm] (h1) {$\disc^{(k)}_{t-1}$};%
  \node[latent, right=of h1,minimum size=1cm] (h2) {$\disc^{(k)}_{t}$};%
  \node[latent, right=of h2,minimum size=1cm] (h3) {$\disc^{(k)}_{t+1}$};%
  
  \node[latent, above=of h1,minimum size=1cm] (z1) {$z^{(k)}_{t-1}$};%
  \node[const, left=of z1,minimum size=1cm] (z0){}; %
  \node[latent, above=of h2,minimum size=1cm] (z2) {$z^{(k)}_{t}$};%
  \node[latent, above=of h3,minimum size=1cm] (z3) {$z^{(k)}_{t+1}$};%
  \node[const, right=of z3,minimum size=1cm] (z4){}; %
  \node[obs, below=of h1,minimum size=1cm]                               (y1) {$y_{t-1}$};%
  \node[obs, below=of h2,minimum size=1cm]                               (y2) {$y_{t}$};%
  \node[obs, below=of h3,minimum size=1cm]                               (y3) {$y_{t+1}$};
  \node[const, left=of y1,minimum size=1cm] (y0){}; %
  \node[const, right=of y3,minimum size=1cm] (y4){}; %
  \node[latent, below=of y2, minimum size=1cm](theta) {$\theta^{(k)}$};
  % Connect the nodes
  \edge {h1} {y1} ; %
  \edge {h2} {y2} ; %
  \edge {h3} {y3} ; %
  \edge {h1} {h2} ; %
  \edge {h2} {h3} ; %
  \edge {h0} {h1} ;
  \edge {h3} {h4};
  \edge {z1} {z2} ; %
  \edge {z2} {z3} ; %
  \edge {z0} {z1} ;
  \edge {z3} {z4};
  %\strike{\edge {y0} {y1}};
  %\strike{\edge {y1} {y2}};
  %\strike{\edge {y2} {y3}};
  %\strike{\edge {y3} {y4}};
  %\edge {y0} {h1};
  %\edge {y1} {h2};
  %\edge {y2} {h3};
  %\edge {y3} {h4};
  %\edge[bend right] {z1} {y1};
  %\edge {z2} {y2};
  %\edge {z3} {y3};
  \draw [->] (z1) to [out=-50,in=60] (y1);
  \draw [->] (z2) to [out=-50,in=60] (y2);
  \draw [->] (z3) to [out=-50,in=60] (y3);
  \edge {theta} {y1};
  \edge {theta} {y2};
  \edge {theta} {y3};

  % Plates
  \plate {p1} {(h1)(h2)(h3)} {$K$} ;
  \plate {p2} {(z1)(z2)(z3)} {$K$} ;
  \plate {p3} {(theta)} {$K$} ;

\end{tikzpicture}
    \caption{\mymodel{}: a factorial switching linear dynamical system for decomposing functional connectivity patterns.}
    \label{fig:our_model}
\end{figure}

The level of activity of the $k$th subnetwork $\theta_k$ at time $t$ is governed by two components: an (approximately) discrete state $\disc_t^{(k)}\in [0,1]$, and a continuous state $z_t^{(k)}\in\mathbb{R}$. The approximately discrete state $\disc_t^{(k)}$ controls whether a subnetwork is currently contributing to the rate, and aids in interpretability of the latent structure. Meanwhile the continuous states $z_t^{(k)}$ control variation in the amplitude of the activity. We include an ``always-on'' discrete background state $\disc_t^{(0)}=1$, and associated continuous state $z_t^{(0)}$, to capture persistent background noise. In the context of MEA recordings, we can think of $\disc_t^{(k)}$ as indicating whether a subnetwork is currently active, and $\exp\left\{z_t^{(k)}\right\}$ as controlling variation in its activity while it is active. The continuous latent variables $z_t^{(k)}$ follow an autoregressive model, parametrized by a time-invariant matrix $A^{(k)}$ and a diagonal noise term $\sigma_p^{(k)2}$. We summarize \mymodel{} in Figure~\ref{fig:our_model}.

At each time $t$, each entry of the $M$-dimensional observation $y_t$ is sampled according to a Poisson distribution parameterized by additive rates.\footnote{We use a Poisson distribution since our application uses spike count data, but alternative choices could be used here.} The $K+1$ components in the additive rates can be seen as activities from $K$ subnetworks or motifs plus a persistent background noise, the structure of which is given by $\theta_k$. In our MEA application, we can think of $\theta_k$ as specifying the relative contributions of each electrode to a subnetwork.

\subsection{Inferring latent states using variational inference}\label{sec:inference}

In order to effectively learn the latent states $\Theta = \{\disc_t, z_t\}_{t=1}^T$ 
we develop an AEVB algorithm inspired by~\citet{becker2019switching}. Our algorithm makes use of a variational distribution $q(\Theta)$  to approximate the posterior of the model defined in Equation~\ref{eqn:our_model}. 

\begin{sloppypar}
    
Our parametrization of $q(\Theta)$ depends on our modeling choices for the RFSM. In the MLP-RFSM, we use a mean-field, auto-encoded representation, $q(\Theta) = \prod_t\prod_kq(\disc_t^{(k)}; y_t) q(z_t^{(k)}; y_t)$, where
\end{sloppypar}

\begin{equation}
    \begin{aligned}
    \alpha_{q,t}^{(k)} =& \text{MLP}_{q, \disc}^{(k)}\left(y_t\right)\\
    \mu_{t}^{(k)}, \sigma_{t}^{(k)} =& \text{MLP}_{q, z}^{(k)}\left(y_t\right)\\
        q(\disc_t^{(k)}) =& \, \text{Concrete}\left(\alpha_{q, t}^{(k)}, \phi\right)\\
        q(z_t^{(k)}) = & \, \text{Normal}\left(\mu_t^{(k)}, \sigma_t^{(k)2}\right)\label{eqn:variational_dist_mlp}
    \end{aligned}
\end{equation}

\begin{sloppypar}
For the RNN-RFSM, where we expect non-trivial relationships between temporally distant observations, we incorporate dependencies of the form $q(\Theta) = \prod_t\prod_k q\left(\disc_t^{(k)};y_t, \disc_{<t}^{(k)}\right)q\left(z_t^{(k)};y_t, z_{<t}^{(k)}\right)$, where the dependency is introduced as follows:
\end{sloppypar}

\begin{equation}
    \begin{aligned}
    \alpha_{q,t}^{(k)}, h_{q, \disc,t}^{(k)} =& \, \text{RNN}_{q, \disc}^{(k)}\left(y_t, h_{q, \disc, t-1}^{(k)}\right)\\
    \left(\mu_{t}^{(k)}, \sigma_{t}^{(k)}\right), h_{q,z,t}^{(k)} =&\, \text{RNN}_{q, z}^{(k)}\left(y_t, h_{q,z,t}^{(k)}\right)\\
        q(\disc_t^{(k)}) =& \, \text{Concrete}\left(\alpha_{q, t}^{(k)}, \phi\right)\\
        q(z_t^{(k)}) = & \, \text{Normal}\left(\mu_t^{(k)}, \sigma_t^{(k)2}\right)\label{eqn:variational_dist_rnn}
    \end{aligned}
\end{equation}

We then optimize the ELBO, following~\citet{aevb}. We treat \{$\sigma_p^{(k)}$\}, $\{\theta_k\}$,  $\{A^{(k)}\}$, and the parameters of the neural networks as parameters to be optimized using stochastic gradient descent. This follows standard practice in AEVB, where distributions over local variables are inferred while global variables are treated as parameters; see \citet{aevb}. The temperature parameter $\phi$ is shared between the prior and variational distributions and is annealed according to the schedule listed in \cite{gong2025supplement}. To encourage sparsity in the inferred latent structure,  we add an L1 penalty $\lambda \sum_t \sum_k \disc_t^{(k)}$ on the learned $\disc_t^{(k)}$ to the ELBO. The per-epoch  computational cost of our model scales as O(TKM).
%Code is available at [add github link] 

\section{Experimental evaluation}\label{sec:results}
In this section, we evaluate \mymodel{} on a series of real and synthetic datasets.   Unless otherwise stated, we use an MLP to model the functions $f_p$, $f_q$ and $f_z$.  We found that, in the absence of clear stereotypy, there was no clear advantage with using a more flexible model such as an RNN. Complete implementation details are found in Appendix A of the Supplementary Material \citep{gong2025supplement}. For experiments on real data, we conducted posterior checks using Pareto Smoothed Importance Sampling~\citep[PSIS,][]{yao2018yes}, which diagnoses whether the proposal distribution is close enough to the true posterior for reliable inference. In all cases we met the proposed stability check ($\hat{k}<0.7$).

In our model, since $\theta_k$ is always multiplied by $e^{z_t^{(k)}}$, the scales of $\theta_k$ and $z_t^{(k)}$ are non-identifiable. In our visualizations, we rescale the inferred $\theta_k$ so that its maximum value is one and adjust the values of $z_t^{(k)}$ accordingly. For the real data, the spike counts exhibit a very heavy tailed distribution. For this reason, when visualizing the raw data and the continuous activities $e^{z_t^{(k)}}$, we truncate our visualization at the 95th percentile of the raw spike count values. When visualizing the latent subgraphs $\theta_k$, we only show subgraphs that are active ($\disc_t^{(k)}>0.5$) in at least 5\% of time points.

\begin{figure}[ht!]
    \centering
    \begin{subfigure}[t]{.3\columnwidth}
    \centering
    \includegraphics[height=1.5in]{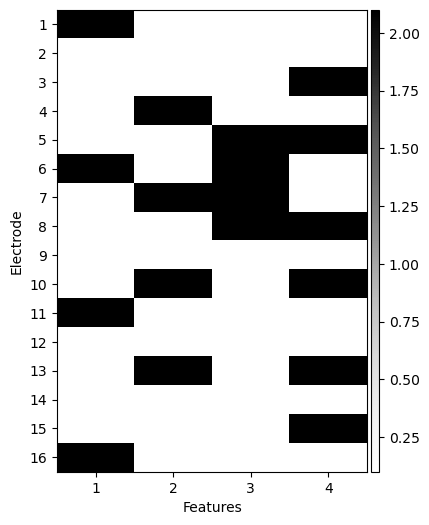}\caption{Latent features $\theta_{1:4}$.}\label{fig:true_features}
    \end{subfigure} ~
    \begin{subfigure}[t]{.33\columnwidth}\centering
    \includegraphics[height=1.5in]{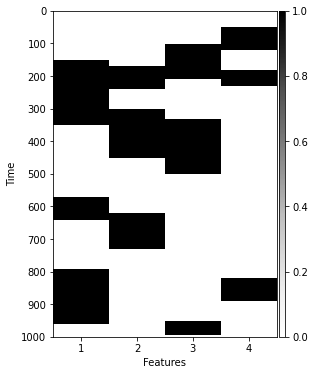}\caption{Feature indicators $\disc_t^{(1:4)}$.} %The highlighted segments at row $k$ indicate feature $k$ is turned on for this time period, revealing how the model partitions the data into different features at different time.}
    \label{fig:true_binary}
    \end{subfigure} ~
    \begin{subfigure}[t]{.3\columnwidth}
    \centering
    \includegraphics[height=1.5in]{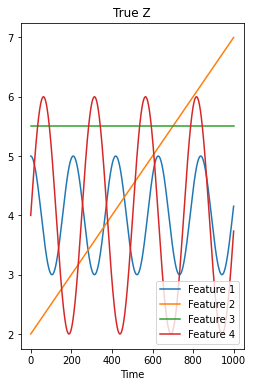}\caption{Latent rates $z_t^{(1:4)}$}.\label{fig:true_continuous}
    \end{subfigure} \\
    
    \begin{subfigure}[t]{.3\columnwidth}
    \centering
    \includegraphics[height=1.5in]{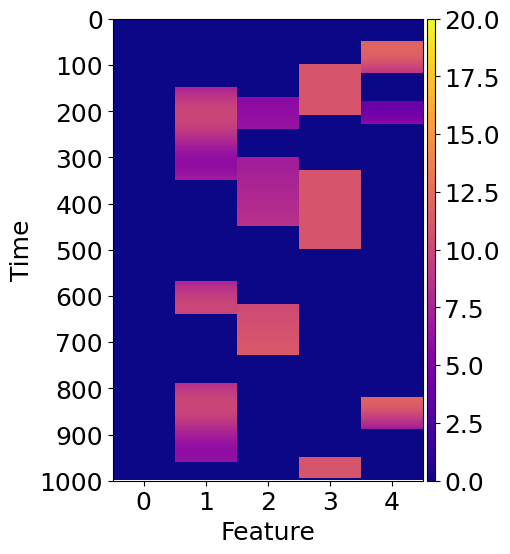}\caption{Combined activities $e^{z_t^{(0)}}, \{\disc_t^{(k)} e^{z_t^{(k)}}\}_{k=1}^4$}\label{fig:true_combined}
    \end{subfigure} ~
    \begin{subfigure}[t]{.33\columnwidth}~
    \centering
    \includegraphics[height=1.5in]{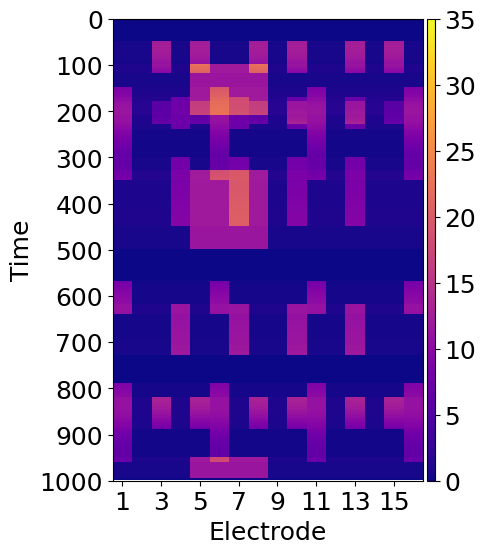}\caption{Combined rate $\sum_{k=1}^4\theta_k\disc_t^{(k)}e^{z_t^{(k)}}$.}\label{fig:true_rate}
    \end{subfigure} ~
    \begin{subfigure}[t]{.3\columnwidth}
    \centering
    \includegraphics[height=1.5in]{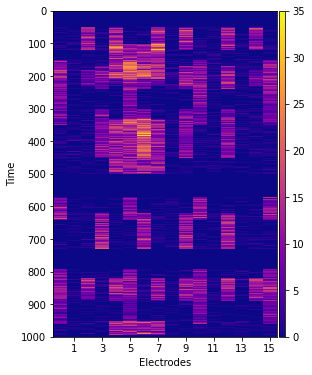}\caption{Observed data $y_t$.}
    \end{subfigure}
        \caption{Synthetic data: Ground truth.}
    \label{fig:sim}
\end{figure}

\subsection{Simulated Data}

\begin{figure}[ht!]
    \centering
    \begin{subfigure}[t]{.24\columnwidth}
    \centering
    \includegraphics[height=1.5in]{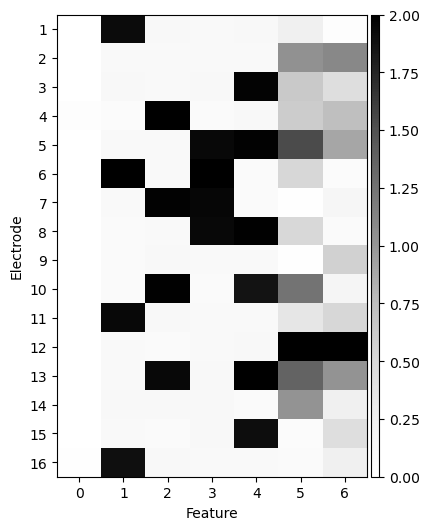}\caption{Latent features $\theta_{0:6}$.}\label{fig:learned_features}
    \end{subfigure}
    \begin{subfigure}[t]{.24\columnwidth}
    \centering
    \includegraphics[height=1.5in]{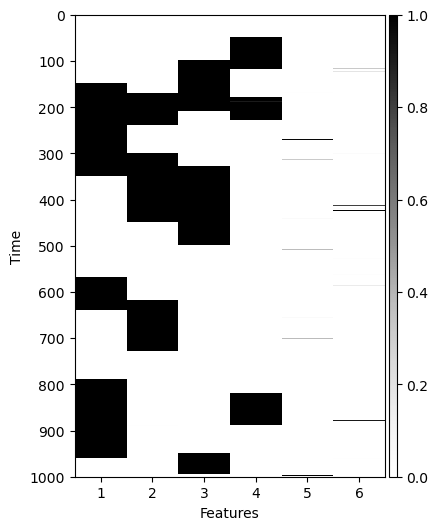}\caption{Feature indicators $\disc_t^{(1:6)}$.}\label{fig:learned_binary}
    \end{subfigure} 
    \begin{subfigure}[t]{.24\columnwidth}
    \centering
    \includegraphics[height=1.5in]{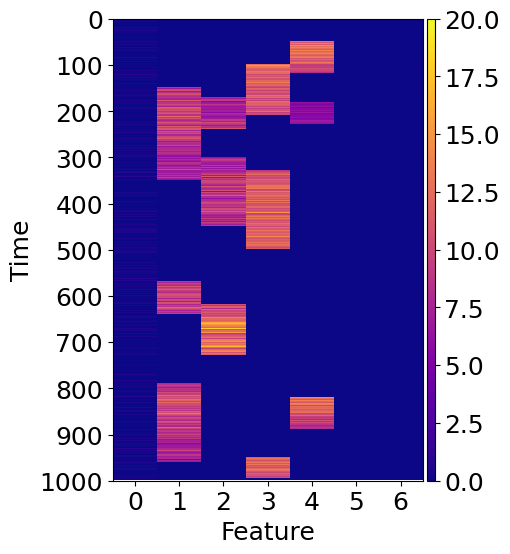}\caption{Combined activities $e^{z_t^{(0)}}$, $\{\disc_t^{(k)} e^{z_t^{(k)}}\}_{k=1}^6$.}\label{fig:learned_combined}
    \end{subfigure} 
    \begin{subfigure}[t]{.24\columnwidth}
    \centering
    \includegraphics[height=1.5in]{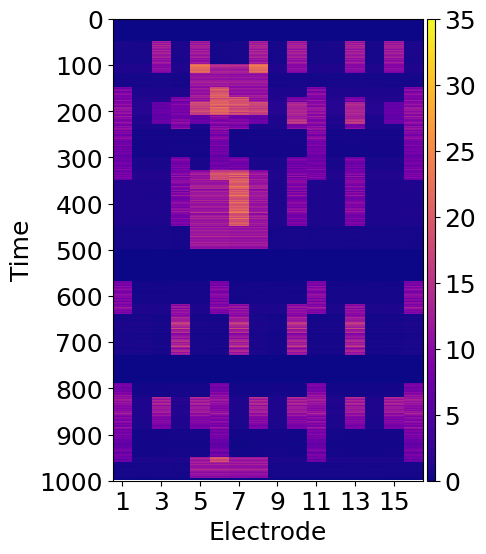}\caption{Combined rate $\theta_0 e^{z_t^{(0)}} + \sum_{k=1}^6\theta_k\disc_t^{(k)}e^{z_t^{(k)}}$.}
    \end{subfigure}
        \caption{Synthetic data: Posterior mean obtained using AEVB. Note that, since the inferred $\disc_t^{(5)}$ and $\disc_t^{(6)}$ are zero everywhere, features $\theta_5$ and $\theta_6$ in Figure~\ref{fig:learned_features} not contribute to the model.}
    \label{fig:sim_result}
\end{figure}
We begin by exploring the performance of \mymodel{} in a synthetic example where the ground truth features are known. % (but \add{generated using sine cosine and linear functions instead of random walks for better visualization purposes}). 
This allows us to explore the ability to recover latent structure and showcase the benefits of using a factorial model over a simpler SLDS. We generate $T=1000$ time steps, with $K=4$ latent, 16-dimensional features $\theta_k$, as shown in Figure~\ref{fig:true_features}. Each feature has a latent continuous trajectory $z_t^{(k)}$ drawn either from a straight line or a sinusoid, as shown in Figure~\ref{fig:true_continuous}. These were combined with manually generated binary trajectories $\disc_t^{(k)}$(Figure~\ref{fig:true_binary}, Figure~\ref{fig:true_combined}) to obtain the rate shown in Figure~\ref{fig:true_rate}. This rate was used to sample random Poisson observations.
 
In Figure~\ref{fig:sim_result}, we show the posterior means obtained via the inference procedure described in Section~\ref{sec:inference}, using a regularization coefficient $\lambda=0.1$ (chosen via cross-validation as described in Appendix A of the Supplementary Material \citep{gong2025supplement}). We generate data with $K=4$ true latent features and for our model with $K=6$ to assess whether it can correctly identify the true number of active features. The results indicate that the Poisson rates are accurately recovered and are appropriately decomposed into the true generating features, demonstrating the model's ability to capture the underlying structure of the data. Even though we set $K=6$, the model correctly identifies the four latent features shown in Figure~\ref{fig:sim}. We note that, while the RFSM uses a continuous-valued indicator $\disc_t^{(k)}$, in practice the values obtained are very close to zero or one (Figure~\ref{fig:learned_binary}); we see similar behavior in all other experiments in this section. 

\begin{figure}[ht!]
    \centering
\includegraphics[width=\columnwidth]{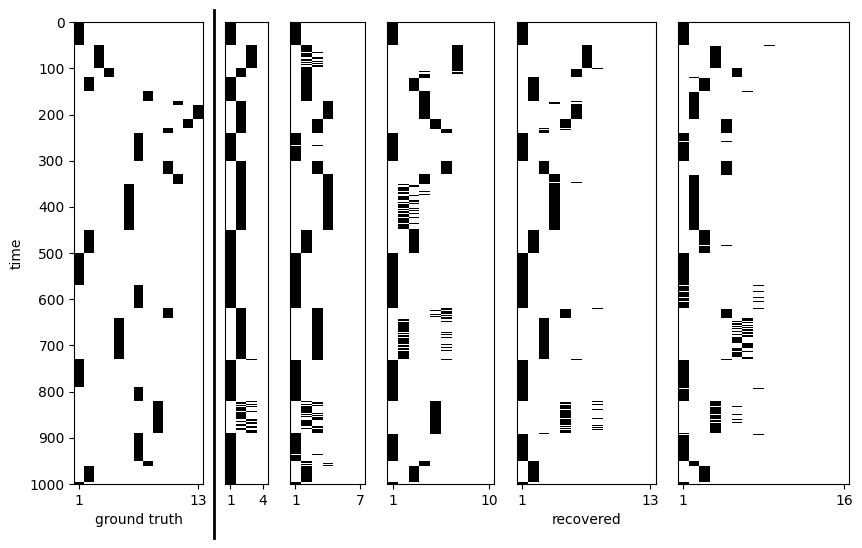}\caption{Latent states obtained using rSLDS. Left hand plot shows the ground truth latent states (corresponding to the thirteen unique combinations of four additive features shown in Figure~\ref{fig:true_binary}). Remaining plots show the results with varying numbers of latent states, from four to sixteen.}\label{fig:slds}
\end{figure}

Next, we consider the latent structure we would find using a non-factorial method. In the left hand side of Figure~\ref{fig:slds}, we have converted the thirteen combinations of latent features apparent in Figure~\ref{fig:true_binary} into thirteen distinct states. This illustrates a key limitation of non-factorial approaches: with K subnetworks, an rSLDS must encode concurrent activity patterns as one of up to $2^K$ discrete states, obscuring which subnetworks are co-active. Already, we see this (true!) latent structure is less interpretable than the additive version in Figure~\ref{fig:true_binary}, since it requires over three times as many features to convey the same information and does not directly reveal which subnetworks are simultaneously active.

To attempt to learn this true, non-additive structure, we run recurrent SLDS (rSLDS), using the \texttt{ssm} package associated with the original paper \citep{pmlr-v54-linderman17a}. We test rSLDS with $K\in\{4, 5,...,16\}$ states  (with the correct number being 13) and 4 latent features,  taking the best ELBO out of ten random seeds. The resulting state assignments are shown in Figure~\ref{fig:slds}. We see that we cannot recover the ground truth. We find that rSLDS tends to miss low-occupancy features and combine multiple features. Even though most of the plots show between two and four main features, these features are not consistent between plots. The results also tend to be noisier than those obtained using \mymodel{}, particularly as the number of features increases. %The results also exhibit more frequent and unnecessary state switching compared to \mymodel{}, particularly as the number of states increases.

In the top row of Table~\ref{tab:allrmse}, we compare the average one-step-ahead predictive RMSE of \mymodel{}, rSLDS, and a linear dynamical system (LDS). Each model was trained on the first 700 observations, and evaluated on the remaining 300. Both rSLDS and LDS are implemented using the \texttt{ssm} package \citep{pmlr-v54-linderman17a}. For rSLDS, we used $K=14$ latent states (selected via cross-validation) and 4 latent features (to match the ground truth). We see that, in addition to yielding more interpretable structure as described previously, \mymodel{} has a notably lower held-out RMSE than the rSLDS and LDS. 

\begin{table}[h!]
  \centering
  \begin{tabular}{l r r r}
    \toprule
    \textbf{Dataset} & \textbf{FSLDS} & \textbf{rSLDS} & \textbf{LDS} \\
    \midrule
    \multicolumn{4}{c}{\textsc{Simulation}} \\
    \cmidrule(lr){1-4}
    Simulation & \textbf{2.4766} & 5.4494 & 5.1786 \\
    \addlinespace[0.5em]
    \multicolumn{4}{c}{\textsc{Hippocampal Data}} \\
    \cmidrule(lr){1-4}
    TC043 DIV14 & \textbf{9.6320} & 10.9240 & 11.5870 \\
    TC165 DIV14 & \textbf{8.2190} & 9.2357 & 11.6800 \\
    \addlinespace[0.5em]
    \multicolumn{4}{c}{\textsc{iPSC Data}} \\
    \cmidrule(lr){1-4}
    P1B2 & 3.8150 & \textbf{3.5248} & 4.9054 \\
    P5A2 & 4.9150 & 6.4342 & \textbf{2.9484} \\
    P3B2 & 3.7388 & \textbf{2.2556} & 2.5836 \\
    \bottomrule
  \end{tabular}
  \caption{RMSE of held-out observations, for simulated data (top); murine hippocampal data (middle) and human iPSC data (bottom).}
  \label{tab:allrmse}
\end{table}

\subsection{Evaluation of microelectrode array spike train data}\label{sec:mea_results}

To evaluate the performance of our algorithm on real-world neuronal connectivity data, we look at two datasets obtained via MEA recordings. For both datasets, we use preprocessed data from ~\citet{sit2024mea}. To obtain this data, spikes (action potentials) were detected from the raw MEA voltage time series, based on the action potential waveform using the continuous wavelet transform method implemented in MEA-NAP, a MATLAB pipeline for MEA network analysis \citep{sit2024mea}.\footnote{\url{https://github.com/SAND-Lab/MEA-NAP}} The voltage timeseries were acquired at a  sampling rate of 12500 Hz (human iPSC-dervied neuronal cultures) and 25000 Hz (murine hippocampal cultures). Action potentials are approximatedly 1ms in duration. For our analysis, we aggregated the spike train data to 1 second time bins.  

\subsubsection{Exploring properties of the inferred structure using murine hippocampal recordings}
\begin{figure}[ht!]
    \centering
    \begin{subfigure}[t]{.23\columnwidth}
    \includegraphics[height=1.8in]{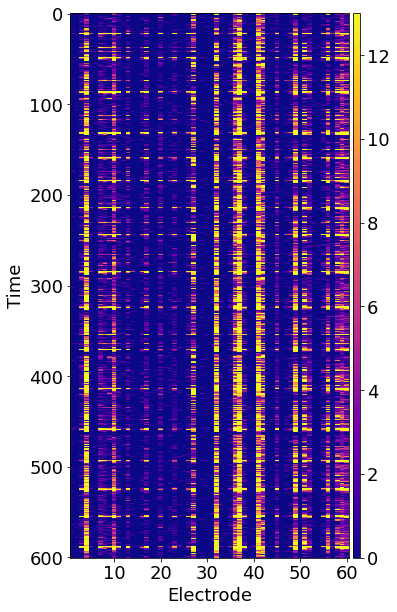}\caption{Binned spike counts.}\label{fig:single_43_raster}
    \end{subfigure} ~
    \begin{subfigure}[t]{.23\columnwidth}
    \includegraphics[height=1.8in]{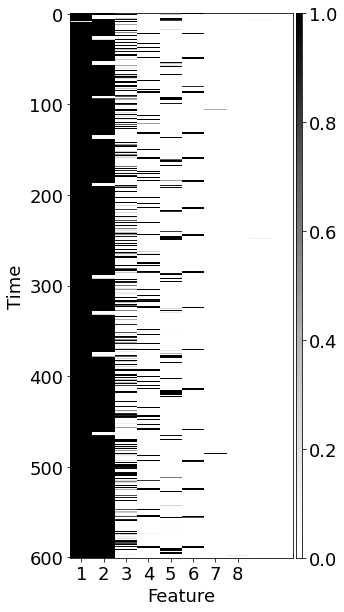}\caption{Feature indicators $\disc_t^{(k)}.$}\label{fig:single_43_learned_h}
    \end{subfigure} ~
    \begin{subfigure}[t]{.23\columnwidth}
    \includegraphics[height=1.8in]{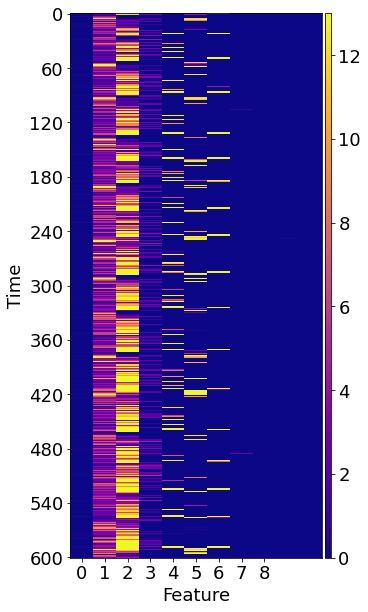}\caption{Combined activities $e^{z_t^{(0)}}, \disc_t^{(k)}e^{z_t^{(k)}}$.}\label{fig:single_43_learned_zh}
    \end{subfigure}     
    \begin{subfigure}[t]{.23\columnwidth}
    \includegraphics[height=1.8in]{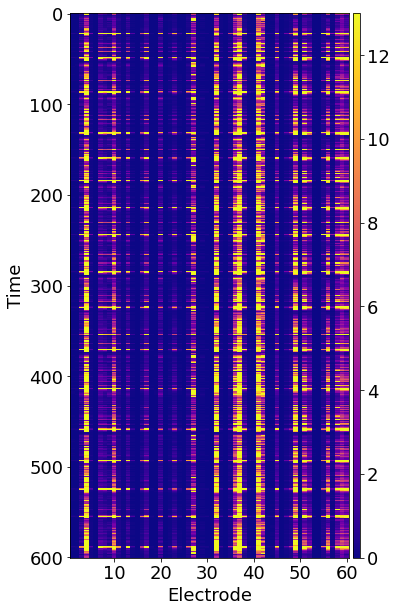}\caption{Combined rate $\sum_k \theta_k \disc_t^{(k)}e^{z_t^{(k)}}$.}\label{fig:single_43_learned_rate}
    \end{subfigure} \\
    \begin{subfigure}[b]{.7\columnwidth}
    \includegraphics[width=\columnwidth]{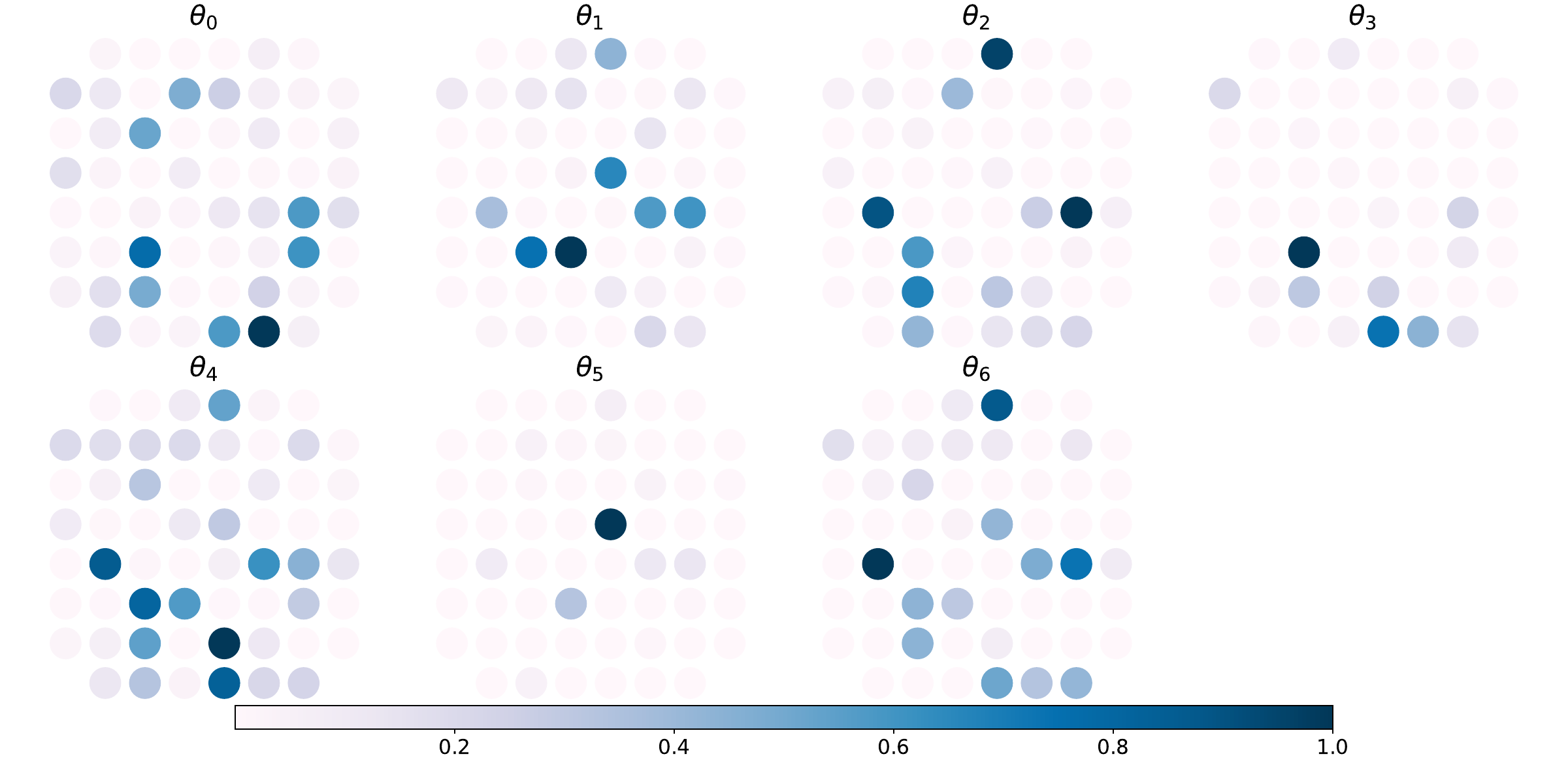}\caption{Latent subgraphs $\theta_k$.}\label{fig:theta_tc043}
    \end{subfigure}
    \caption{Murine hippocampal data (recording TC043 DIV14). Figure~\ref{fig:single_43_raster} shows the spike counts (color bar, truncated to the 95th percentile value) from a 10-minute MEA recording of a days-in-vitro (DIV) 14 primary hippocampal culture. The activity observed in each electrode (columns) is shown over time (y-axis unit is seconds). Spike count binned by 1 second intervals. Figure~\ref{fig:single_43_learned_h} shows the binary feature indicators $\disc_t^{(k)}$, indicating which of the latent features are active at each time point. Figure~\ref{fig:single_43_learned_zh} shows the expression level of each of the latent features $\theta_k$. Figure~\ref{fig:single_43_learned_rate} shows the inferred firing rate. Figure~\ref{fig:theta_tc043} shows the relative contribution of each electrode, in the spatial arrangement of the MEA, for a subset of the latent states $\theta_k$, including the background (subnetwork 0) and most active subnetworks (1-4). The relative contribution (node color) is scaled for each subnetwork. The predictive RMSE of the last 50 seconds is 9.632.}\label{fig:mouse_single_43}
\end{figure}

We first consider qualitative analyses based on MEA recordings from two 2D murine hippocampal cultures (TC043 DIV14 and TC165 DIV14), first published in \citep{schroeter2015emergence}, using the spike train data extracted by \citet{sit2024mea}. Results for TC165 are included in Appendix C.1 of the Supplementary Material \citep{gong2025supplement}. We visualize the neuronal activity from these two recordings in Figures~\ref{fig:single_43_raster} and Figure 1a in the Supplementary Material. Since we see clear stereotypy that would not be captured using a Markovian switching model, we choose to use RNNs to model $f_p$, $f_q$ and $f_z$.
\footnote{Fourier analysis suggests TC043 DIV14 and TC165 DIV14 have an average period of 16.9 seconds and 33.5 seconds respectively.} We used a third recording, from culture TC050 DIV21, as a validation set to select the regularization parameter $\lambda$ as described in Appendix A in the Supplementary Material \citep{gong2025supplement}, yielding a value of $\lambda=0.1$.

The remaining subplots in Figure~\ref{fig:mouse_single_43} (and Figure~1 in the Supplementary Material) show the latent structure obtained using \mymodel{}. In each case, we see the activity decomposed into distinct subgraphs (Figure~\ref{fig:theta_tc043} and Figure~1e in the Supplementary Material). Some features are active throughout most of the recording, but with varying levels of activity (e.g., features 1 and 2 in Figure~\ref{fig:mouse_single_43}). Others are active in short bursts, such as features 5 and 6 in Figure~\ref{fig:mouse_single_43}. Posterior predictive checks using PSIS yielded $\hat{k}=0.51$ for TC043 DIV14 and $\hat{k}=0.34$ for TC165 DIV14, suggesting approximate convergence.

We next look at quantitative performance of FSLDS. In the middle row of Table~\ref{tab:allrmse}, we calculate average one-step-ahead predictive RMSE for the two cultures described above, using FSLDS, rSLDS, and LDS. In each case, we trained on the first 550 observations, and evaluated on the last 50. For rSLDS we select the number of latent states and latent features using held-out cross-validation on culture TC050 DIV21. In both cases, we see better predictive performance using FSLDS, indicating that we do not sacrifice predictive performance in favor of interpretability.

\iffalse
\begin{table}[h!]
  \centering
  \begin{tabular}{c|c|c|c}
    \hline
    &FSLDS & rSLDS & LDS \\
    \hline
    TC043 DIV14 & 9.632 & 10.923985 & 11.587 \\
    TC165 DIV14 & 8.219 & 9.235658 & 11.680 \\
    \hline
  \end{tabular}
  \caption{Hippocampal data RMSE}
  \label{tab:mousermse}
\end{table}
\fi

One concern might be that the number of inferred latent features will grow with sequence length regardless of the true latent dimensionality. In Appendix B in the Supplementary Material \citep{gong2025supplement}, we demonstrate that this is not the case: We show that the structure uncovered in Figure~\ref{fig:mouse_single_43} (and Figure 1 in the Supplementary Material) persists if we vary the recording length.

\FloatBarrier

\begin{figure}[ht!]
    \centering
    \includegraphics[width=0.8\columnwidth]{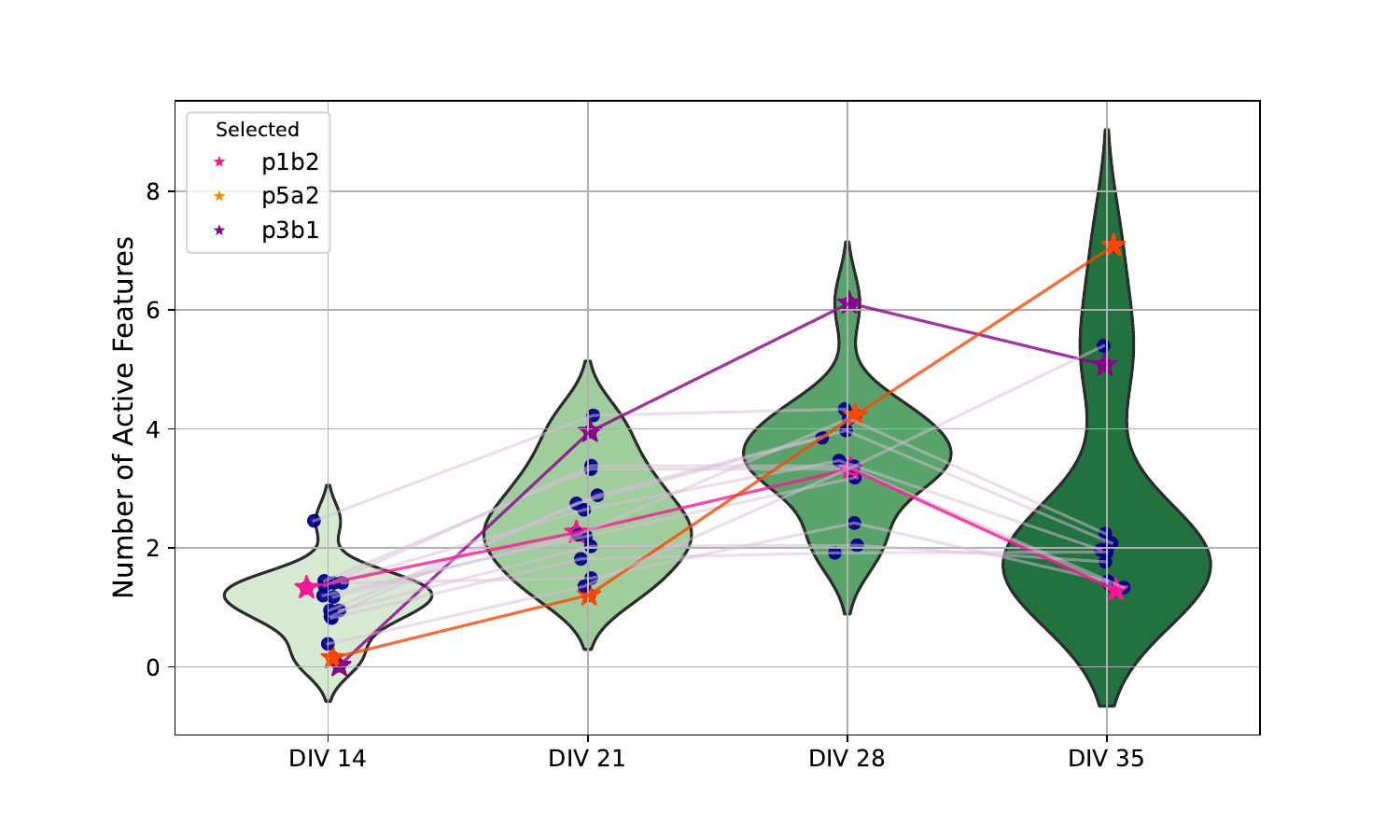}
    \caption{Number of activated features across weeks in culture. Violin plots with scatter plots of number of active features for each culture (circles or stars, n=15 cultures) at four developmental time points days-in-vitro (DIV) 14, 21, 28, and 35. Lines show developmental trajectory of number of latent features within individual cultures. Three examples are highlighted with stars and thicker colored lines: P1B2 (pink, see also Figure~\ref{fig:p1b2}), P5A2 (orange, see also Figure~4 in the Supplementary Material \citep{gong2025supplement}), and P3B1 (purple, see  also Figure~5 in the Supplementary Material).}
    \label{fig:violin}
\end{figure}

\subsubsection{Exploring evolution of connectivity over longer time periods}
Next, we consider a dataset of MEA recordings from human induced pluripotent stem cell (iPSC)-derived neurogenin-2 (NGN2) cortical neurons cultured on 64-electrode MEAs (n=15 cultures), first analyzed in \citet{sit2024mea}. Each culture was recorded for a period of ten minutes once per week, over a period of four weeks, at days-in-vitro (DIV)14, 21, 28 and 35 for $n=15$ cortical cultures. Spike counts were binned into one-second intervals. To investigate subnetwork development in vitro, we concatenated the data from all four weeks for each culture. We randomly select one dataset, P1A2, as the validation set to choose the penalty $\lambda$, as described in Appendix A in the Supplementary Material \citep{gong2025supplement}. Here $\lambda$ is selected as 0.02, which is associated with the lowest reconstruction error.%Sample raster plots are shown in Figure~\ref{fig:data}.

We expect the level of functional connectivity to increase as the neuronal networks mature, reflected in both the number of active subnetworks and the overall complexity of the resulting network. We expect changes in the neuronal activity at the end of each recordings at the different days-in-vitro (since they correspond to temporal discontinuities) and changes in the latent structure within each time period. We set a maximum of 10 latent features and ran our variational algorithm using thirty-five different seeds, selecting the result with the highest ELBO. 

\begin{figure}[t]
\centering
    \begin{subfigure}[t]
    {0.45\columnwidth}
    \centering
\includegraphics[width=.8\columnwidth]{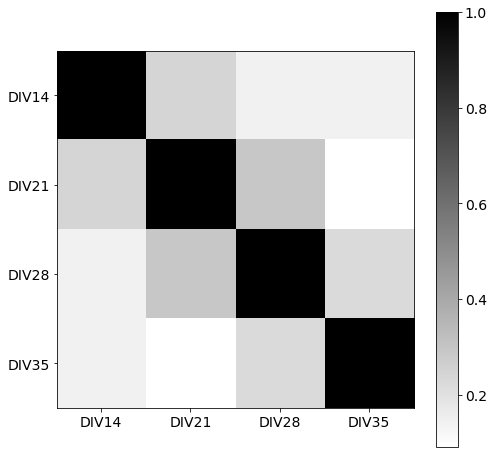}\caption{Mean cosine similarities.}
    \end{subfigure}~
\begin{subfigure}[t]{0.45\columnwidth}
\centering
\includegraphics[width=.8\columnwidth]{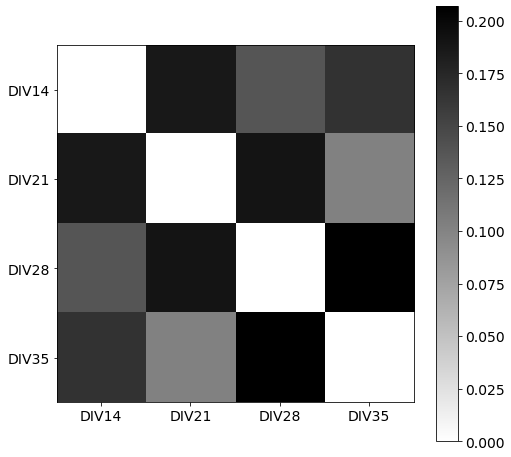}\caption{Standard deviation of cosine similarities.}
    \end{subfigure}
    \caption{Mean and standard deviation of the cosine similarities between the vectors $\left(\disc_t^{(1)},\dots, \disc_t^{(K)}\right)$ for each time period (DIV14, 21, 28 and 35) across the concatenated recordings.}\label{fig:cosine}
\end{figure}

\begin{figure}[ht]
    \centering
    \begin{subfigure}[t]{.23\columnwidth}
\includegraphics[height=1.8in]{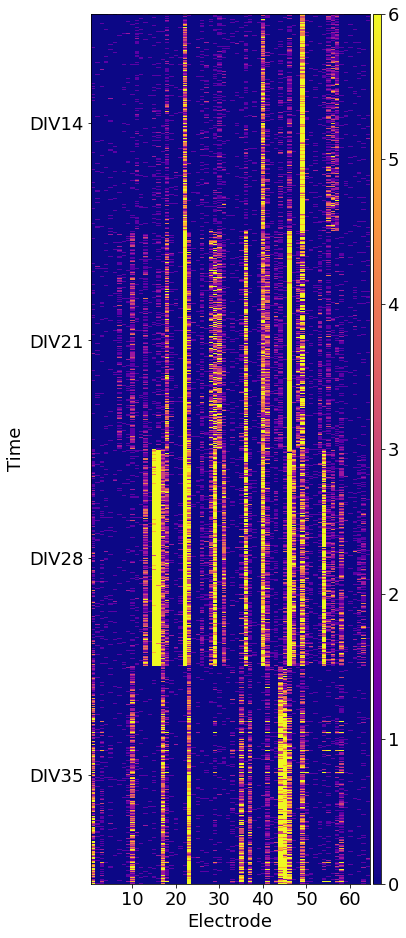}\caption{Binned spike counts.}\label{fig:raster1}
    \end{subfigure}
    \begin{subfigure}[t]{.23\columnwidth}
    \includegraphics[height=1.8in]{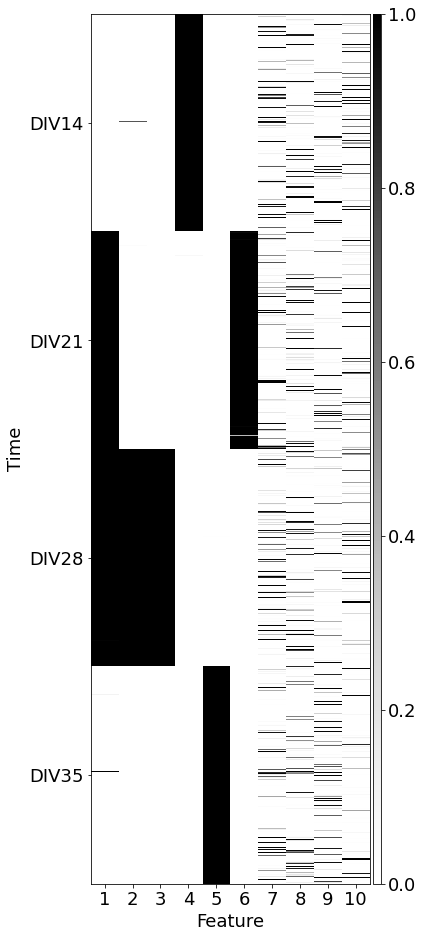}\caption{Feature indicators $\disc_t^{(k)}$.}\label{fig:p1b2_learned_h}
    \end{subfigure} ~
    \begin{subfigure}[t]{.23\columnwidth}
    \includegraphics[height=1.8in]{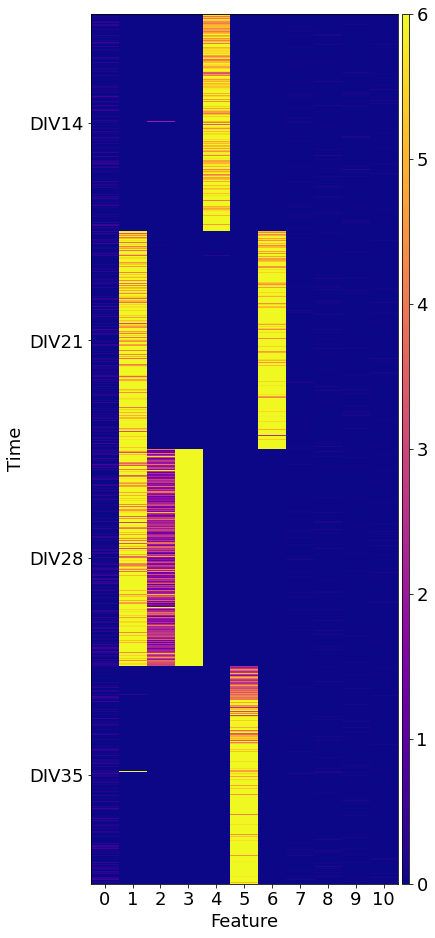}\caption{Combined activities $e^{z_t^{(0)}}, \disc_t^{(k)}e^{z_t^{(k)}}$.}\label{fig:p1b2hz}
    \end{subfigure} ~
    \begin{subfigure}[t]{.23\columnwidth}
    \includegraphics[height=1.8in]{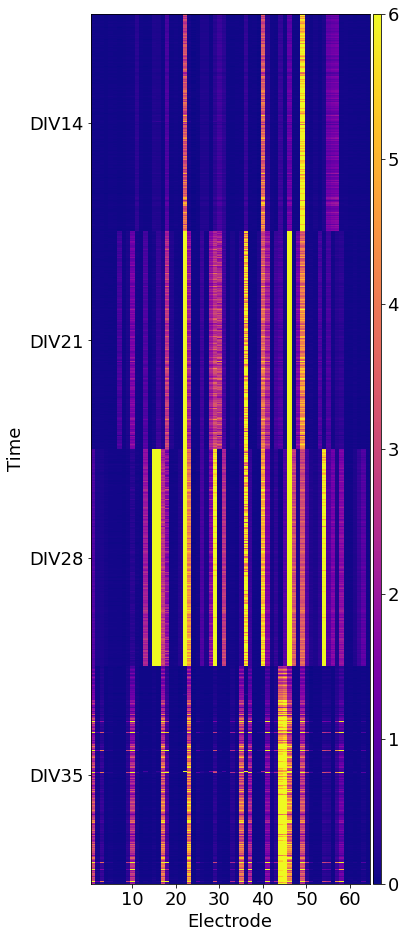}\caption{ Combined rate $\sum_k \theta_k\disc_t^{(k)}e^{z_t^{(k)}}$.}\label{fig:p1b2rate}
    \end{subfigure} ~

    \begin{subfigure}[b]{.9\columnwidth}
    \includegraphics[width=\columnwidth]{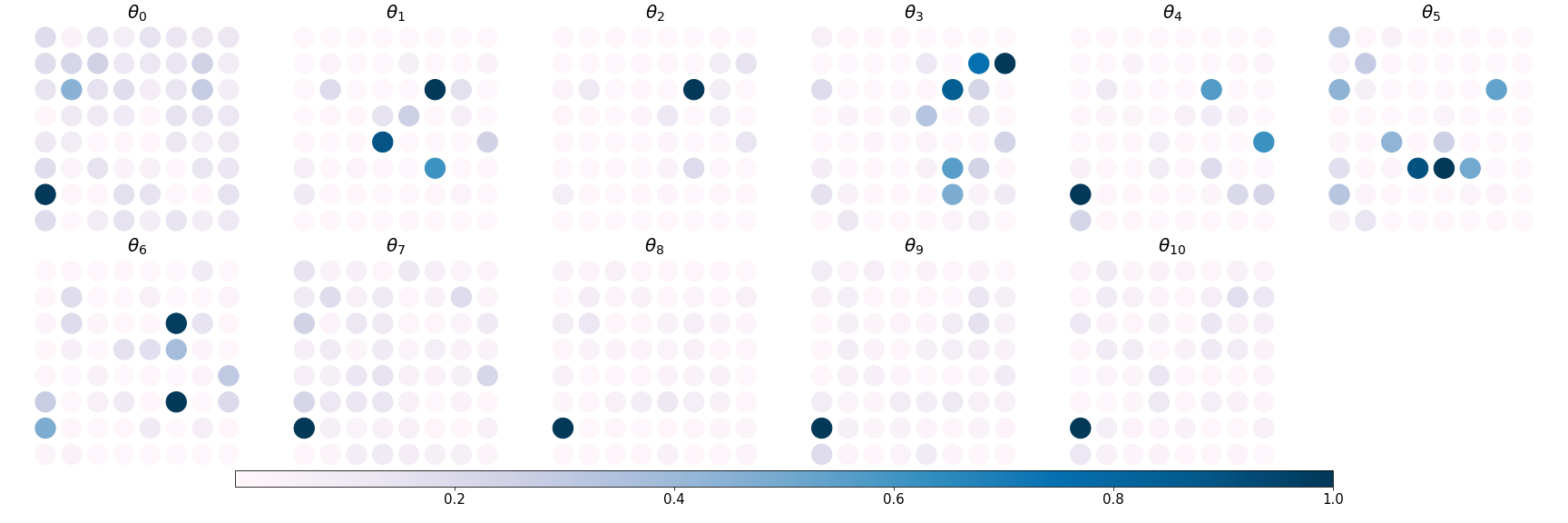}\caption{Latent subgraphs $\theta_k$.}\label{fig:theta_p1b2}
    \end{subfigure} ~
    \begin{subfigure}[b]{.9\columnwidth}
    \includegraphics[width=\columnwidth]{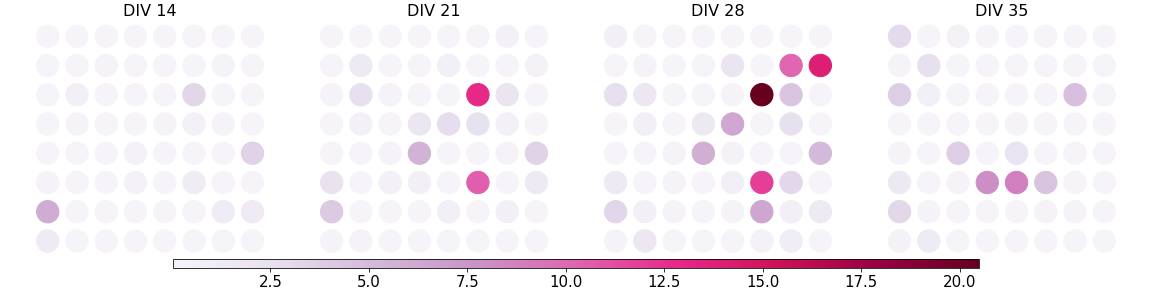}\caption{Mean network activity.}\label{fig:p1b2weeklysub}
    \end{subfigure} ~
        \caption{Development of human iPSC-derived NGN2 cortical neurons in culture P1B2. Figure~\ref{fig:raster1} shows the raster plot from the concatenated spike train data from days-in-vitro (DIV) 14, 21, 28, and 35 with the spike count (binned in 1 second intervals) for the 64 electrodes (columns) over the four 10-minute recordings (x-axis unit is seconds). Figure~\ref{fig:p1b2_learned_h} shows the latent features (columns) for the 10-minute MEA recordings from the four developmental time points (rows). Color bar represents the probability of the feature being on (yellow) or off (purple). Figure~\ref{fig:p1b2hz} shows the combined activity (color bar) revealing the temporal dynamics within the on and off switching of the latent states within each MEA recording. Lighter colors (pink to yellow) show dominant features in contrast to features that are off (dark blue). Figure~\ref{fig:p1b2rate} shows the inferred firing rate (color bar). Figure~\ref{fig:theta_p1b2} shows heatmaps of the relative contribution (color) of individual electrodes (circles) to the subgraphs in the spatial arrangement of the MEA for subnetworks 0-7. Figure~\ref{fig:p1b2weeklysub} shows heatmaps  in the spatial arrangement of the MEA of the mean spike count (per second) per electrode (color) at the four developmental time points. PSIS coefficient is -0.09<0.7.}
    \label{fig:p1b2}
\end{figure}

In Figure~\ref{fig:violin}, we show the average number of active features (or subnetworks) for each week, for each recording. In almost all cultures, we see an increase in the number of active features across the four weeks, suggesting a developmental process. This coincides with an increase in the number of active electrodes and the number of action potentials detected by these electrodes \citep{sit2024mea}. However, this pattern diverges at DIV 35, where a bimodal behaviour in the dataset is observed.  While a few cultures continue to show a similar number (or increase) in the number of subgraphs at DIV 35 compared to DIV 28 (e.g., P5A2, shown in Figure~5 in Appendix C.2 in the Supplementary Material \citep{gong2025supplement}), in other cultures, this number decreases (e.g., Figure~\ref{fig:p1b2}).  These likely represent more mature cultures in which the increase in the degree of connectivity has led to a consolidation of subgraphs between DIV 28-35 \citep[see][]{sit2024mea}.

In Figure~\ref{fig:cosine}, we look at the overlap between the connectivity across different weeks. We average $\disc_t^{(k)}$ over each week and concatenate to get a $K$-dimensional vector. We then calculate the cosine similarities between these vectors. High values indicate high overlap between the average connectivity patterns; low values indicate little overlap. We see moderate overlap adjacent weeks, but low overlap over longer timescales.

Next, we look at the recovered structure for three cultures in more detail. We selected these recordings because they are representative of key types of behavior shown in Figure~\ref{fig:violin}. The developmental trajectory of these cultures can be identified in Figure~\ref{fig:violin} by the color of the stars: P1B2 (pink), P5A2 (orange), and P3B1  (purple).

\afterpage{\FloatBarrier}
In the first culture, P1B2 (Figure~\ref{fig:p1b2}), we see an initial increase in the number of subnetworks from DIV 14-28, followed by a mild consolidation of the network structure into new activity patterns at DIV 35. This illustrates the ability of the model to detect biologically relevant changes in subnetworks as the neuronal networks mature.  The model detects seven different features (Figure~\ref{fig:p1b2_learned_h}--\ref{fig:p1b2hz}), and the latent subgraphs reveal different combinations of electrodes detecting activity in these subnetworks (Figure~\ref{fig:theta_p1b2}). From DIV 14 to DIV 28, we observe an increasing number of different features that are turned on (from 1 to 3). In addition, we see one feature introduced at DIV 21 remains active during DIV 28. However, at DIV 35, one completely different feature is activated.  These findings are consistent with an increase in network activity (Figure~\ref{fig:p1b2weeklysub}) and density between DIV14-28, illustrated in \citet{sit2024mea}, and an increase in the mean nodal participation coefficient between DIV 28-35. When considering the spatial arrangement of the subnetworks (Figure~\ref{fig:theta_p1b2}), we also observe a shift in participating nodes. During the first three weeks, the active nodes are more likely to be located in the middle or right side of the MEA, while at DIV35, the activity has shifted to the left.  This may indicate that some subnetworks may increase or decrease in dominance in the overall network activity as the cultures mature.

Figures for two additional cultures are shown in Appendix C.2 in the Supplementary Material \citep{gong2025supplement}. The second culture, P5A2 (Figure~4 in the Supplementary Material), clearly demonstrates an increase in network complexity from DIV 14-35 (Figures~4b--4c in the Supplementary Material), without the consolidation behavior seen in P1B2 at DIV 35. With no active features observed at DIV14, subnetwork 0 (Figure~4e in the Supplementary Material) mainly comes from the baseline noise. As the neuronal networks mature, more features are activated, forming more complex subnetworks (Figures 4c--4e in the Supplementary Material). We hypothesize that P5A2 may be following a similar developmental trajectory to P1B2, but with a slower start (as evidenced by the lack of correlated activity at DIV 14). As with P1B2, we observe the overall network structure becoming more complex as additional subnetworks are activated.

The third example, P3B1 (Figure~5 of the Supplementary Material  \citep{gong2025supplement}), shows a growth and then stabilization of network complexity.  At DIV 14, no active features are observed, and most of the activity shown in subnetwork 0 is due to the low-level background process. From DIV 21-35, there is a developmental increase in the number of active features. We see high overlap in the active subnetworks: subnetworks $\theta_1$ and $\theta_2$ persist from DIV21-28 and $\theta_5$ persists from DIV 28 -- DIV 35. Additionally, three  subnetworks dominated by a single electrode ($\theta_6$, $\theta_7$ and $\theta_{10}$) persist from DIV 14 to DIV 35.

In the bottom row of Table~\ref{tab:allrmse}, we compare the average one-step-ahead predictive RMSE of FSLDS with rSLDS and LDS, on the three cultures explored in this section. For rSLDS, we select the number of latent states and latent features using held-out cross-validation on a fourth culture P1A2. In this case, FSLDS does not perform better than rSLDS and LDS in terms of RMSE. However, this is perhaps not surprising: the held-out test set only contains timepoints in DIV35. Unlike the murine hippocampal data, in the iPSC cultures we do not see significant changes in the latent features \textit{within} a time period, so there is no predictive benefit in modeling such transitions. However, this is offset by the fact that FSLDS does give us insights into the switching behavior \textit{between} time periods, allowing insights about neuronal development that are not directly accessible via the existing LDS/rSLDS approaches.

\section*{Conclusion}

We proposed a new factorial switching dynamical system model to detect changes in brain networks. This model allows concurrent changes in subnetworks, without assuming unrealistic abrupt changes in the whole network. We pair it with a scalable inference algorithm enabling fast computation. The model performs well on MEA data from murine primary hippocampal cultures and from human iPSC-derived cortical cultures.  It captures a developmental increase in the number of subnetworks and the likelihood of these subnetworks being on during the MEA recordings. Although LDS/SLDS occasionally achieves higher prediction accuracy, this is not unexpected: \mymodel{} assumes that dynamics modes can be expressed as weighted combinations of learned basis states, so with sufficient data both approaches should capture similar structure. However, \mymodel{} provides a more interpretable parameterization by explicitly decomposing network activity into constituent subnetworks and their temporal dynamics. Several natural extensions could be pursued in future research. For example, in this paper, we use a fixed number of features $K=10$ (in practice, we found ten features sufficient to model the real-world datasets explored here) and use a regularization parameter $\lambda$ to shrink the number of active features. An interesting future line of work would be to learn the number of features, either by imposing a prior on $K$ or by using a nonparametric Bayesian approach.

This model provides an important tool for neurobiologists studying brain development and brain disorders with MEA recordings. It extends the tools beyond pair-wise comparisons to detect functional connectivity and identifies patterns of activity from subnetworks of different sizes that can vary on the scale of seconds to minutes during the MEA recording.  This will enable neurobiologists to investigate differences in network complexity, capacity for information sharing, and stability versus flexibility of state switching in subnetworks.  As the method uses spike train data, it may also be useful for analysis of data from other multielectrode recordings (e.g., Neuropixel probes), which may be performed in vivo. This method could also be used with MEA recordings to compare the effect of drugs on modulating network states in vitro or the effect of genetic mutations on the number of states and frequency of state switching. Thus, this approach may inform future mechanistic and therapeutic investigations in human-derived or murine models of neurological disorders. 

\section*{Supplementary Materials}

The supplementary materials include inference implementation details, learned structure evaluation given growing sequence length and further details on real data analysis.

%\newpage
\bibliography{references}
\bibliographystyle{apalike}

\end{document}